\newcommand{\be}{\begin{equation}}
\newcommand{\ee}{\end{equation}}
\newcommand{\bea}{\begin{eqnarray}}
\newcommand{\eea}{\end{eqnarray}}
\def\pslash{{\cal P}{\hbox{\kern-6pt $\slash$}}}

\long\def\comment#1{}

\documentstyle[prl,aps,floats,twocolumn]{revtex}

\input epsf
\begin{document}
\draft
\twocolumn[\hsize\textwidth\columnwidth\hsize\csname @twocolumnfalse\endcsname
\title{ The dearth of halo dwarf galaxies: is there power on short scales?}
\author{Marc Kamionkowski\cite{marcmail}}
\address{California Institute of Technology, Mail Code 130-33,
Pasadena, CA 91125}
\author{Andrew R. Liddle}
\address{Astrophysics Group, Blackett Laboratory, Imperial
College, Prince Consort Road, London, SW7 2BZ, United Kingdom\\
Isaac Newton Institute, University of Cambridge, Cambridge CB3 0EH, United 
Kingdom}
\date{\today}
\maketitle

\begin{abstract}
$N$-body simulations of structure formation with scale-invariant
primordial perturbations show significantly more
virialized objects of dwarf-galaxy mass in a typical
galactic halo than are observed around 
the Milky Way.  We show that the dearth of observed
dwarf galaxies could be explained by a dramatic downturn in the
power spectrum at small distance scales.  This suppression of
small-scale power might also help mitigate the disagreement between
cuspy simulated halos and smooth observed halos, while remaining
consistent with Lyman-alpha-forest constraints on small-scale
power.  Such a spectrum could arise in inflationary models with
broken scale invariance.
\end{abstract}

\pacs{\hfill CALT-68-2249, astro-ph/9911103}
]

The broad-brush picture painted by the inflation-inspired
hierarchical-clustering paradigm accounts for the smoothness of
the cosmic microwave background (CMB), its tiny temperature
fluctuations, the flatness of the Universe, and the observed
distribution of galaxies.  However, finer inspection yields some 
possible---and possibly troubling---discrepancies between the
models and observations that still need to be ironed out.  One
of these is the dearth of substructure in galaxy halos. 

Recent
high-resolution $N$-body simulations \cite{KlyKraVal99,Mooetal99}
have confirmed earlier analytic arguments \cite{KauWhiGui93}
that suggested that hierarchical-clustering models should
produce far more dwarf galaxies around the Milky Way than are
observed.
There are only 11 dwarf galaxies with internal velocity
dispersions greater than 10 km~sec$^{-1}$ within the virial radius of 
the Milky Way halo.  However, numerical simulations of structure 
formation in a hierarchical model indicate that a halo of the
Milky Way's mass and circular speed should contain roughly an
order of magnitude more dwarf galaxies.  These theoretical
results are robust to changes in the values of cosmological
parameters or in the tilt of the primordial spectrum of
perturbations.  

One might at first be tempted to dismiss this discrepancy 
between theory and observations as a consequence of some nasty
astrophysics.  After all, the simulations consider only
gravitational interactions and identify only virialized dark
halos, while the real Universe is filled with gas, and dwarf
galaxies are identified by their visible matter.  So, for
example, one might guess that the halos are there but remain
invisible because the gas has been expelled by an early
generation of supernovae.  However, even generous estimates of
the efficiency of supernova-driven winds fall short of
explaining the absence of luminous dwarf galaxies \cite{MacFer98}.
Further, even if the baryonic matter could somehow be driven out of the
mini-halos or kept dark, we would still have trouble explaining
how a spiral disk could have formed in the strongly fluctuating
potential of such a clumpy halo.  See
Refs. \cite{KlyKraVal99,Mooetal99,SpeSte99} for more detailed
reviews of such arguments.

In the absence of any prosaic astrophysical mechanism, it is
natural to think of more exotic explanations. One strategy is to modify the 
nature of the dark matter, in order to prevent low-mass halos
{}from forming. One 
option is that the dark matter is warm, for example a neutrino of mass
around a keV, which would suppress the formation of
small-scale structure by free-streaming out of potential wells
\cite{SchSil88}.  However, such a neutrino would diminish power
on scales $2$ to $12\,h^{-1}$ Mpc to a degree that may conflict with 
the power inferred from the Lyman-alpha forest
\cite{SpeSte99,Croetal98}.  Another possibility \cite{SpeSte99}
is that the dark-matter particles interact strongly with
each other, but not with ordinary matter
\cite{CarMacHal92}.  However, the properties
required of this particle (elastic-scattering cross-sections of
order $10^{-25}\,{\rm cm}^2$) are almost inconceivable in the
predominant paradigms of  weakly-interacting massive particles
\cite{JunKamGri96} and axions \cite{Tur90}.

Here we consider an alternative strategy: broken scale
invariance in the primordial power spectrum, arising from some feature in the 
inflaton 
potential.  
In slow-roll inflation, the amplitude of density perturbations
on some comoving scale is proportional to $V^{3/2}/V'$, where
$V$ is the inflaton potential, and $V'$ its derivative, when the 
comoving scale under consideration exits the horizon.  Suppose
that $V'$ is initially very small and that there is then a
discontinuity in the second derivative, $V''$.  The slope $V'$
will then jump, and the density-perturbation amplitude will
drop steeply.  In this way, the density-perturbation amplitude
on some suitably small scale can be suppressed relative to the
power on some larger scale.  Such an idea has been invoked
as a possible explanation of a claimed break in the power
spectrum at  $\sim100$ Mpc scales \cite{break,LPS} (and in an
attempt produce non-Gaussian features \cite{limin}).  Here, we
use this idea to account for the lack of halo substructure.
Below, we show how a suppression of small-scale power can affect 
the dwarf-galaxy abundance and illustrate a particular
inflationary model.  We also argue briefly that the absence of
small-scale power may help explain the discrepancy between
simulated cuspy halos \cite{NFW} and observed smooth halos.

\begin{figure}[t!]  
\epsfxsize=3.3 in \epsfbox{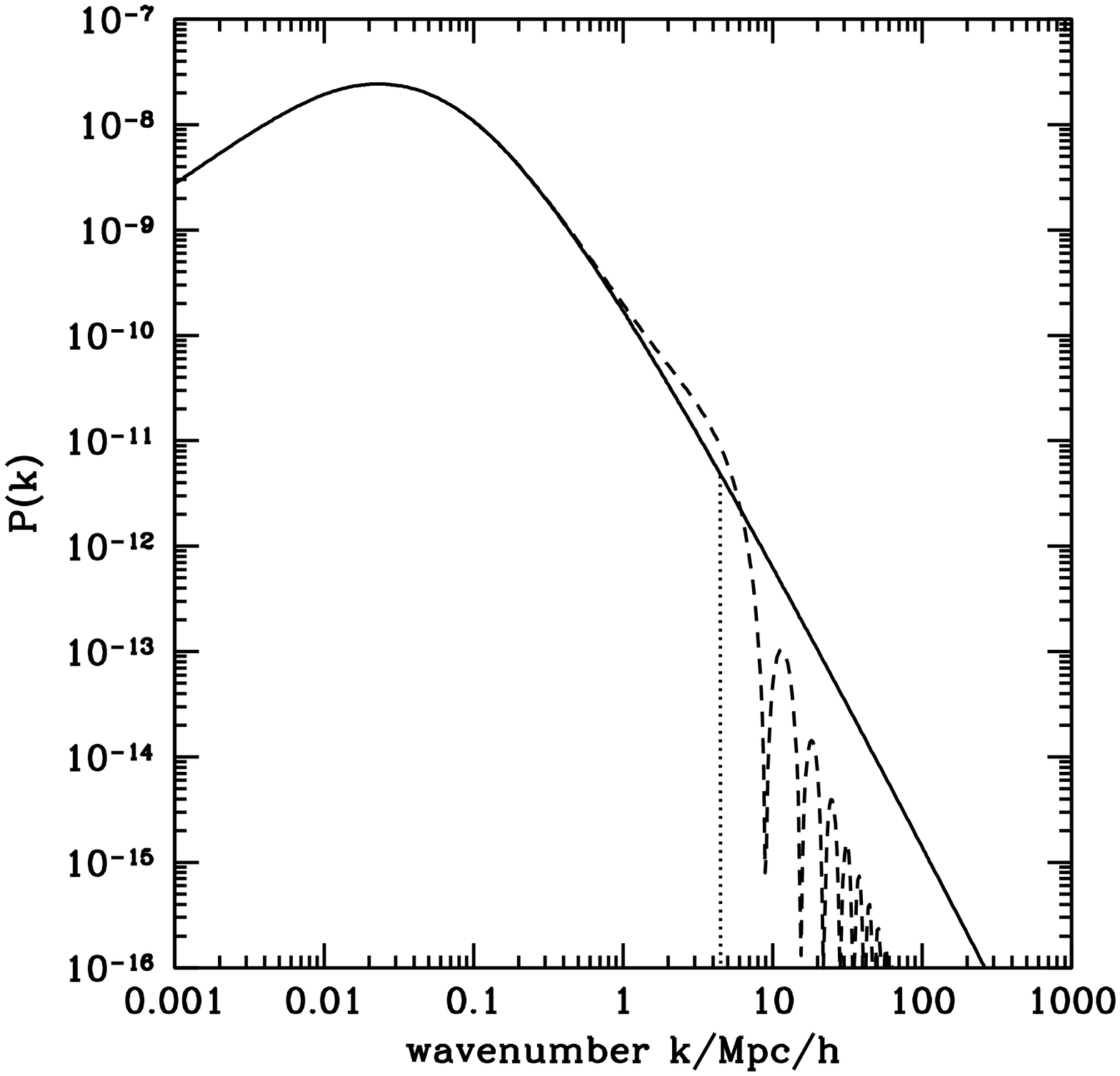}  
\epsfxsize=3.3 in \epsfbox{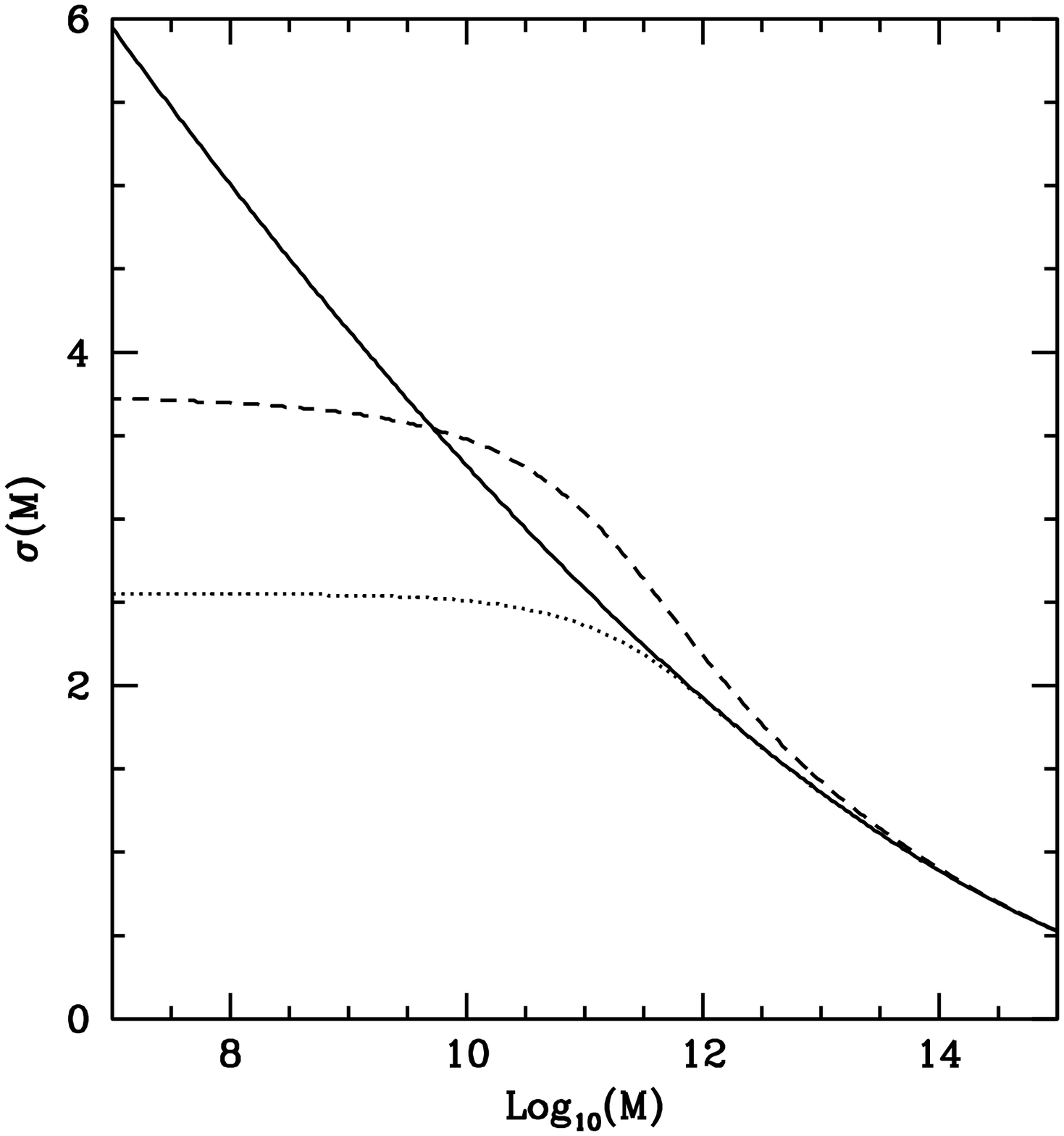}  
\caption{The upper panel shows the power spectrum for $\Gamma=0.25$ CDM (solid 
curve), for a model in which the power spectrum is arbitrarily cut off at 
$k=4.5\,h$ Mpc 
(dotted curve), and the broken-scale-invariance inflation model
(dashed curve).
The lower panel shows the rms mass fluctuation as a function of 
the enclosed mean mass $M$ for these three models.}
\label{fig:powerspectrum}
\end{figure}

For simplicity and clarity of presentation, we restrict our discussion here to
an Einstein-de Sitter (EdS) model, but everything can readily be generalized to
a low-density Universe.  Although we focus on an EdS cosmogony, we use a
cold-dark-matter (CDM) power spectrum \cite{cdmpowerspectrum} with $\Gamma=0.25$
(e.g., as in a $\tau$CDM model \cite{tauCDM}) to ensure that observational
constraints from the shape and amplitude of the galaxy power spectrum, the
cluster number density, and the COBE normalization are satisfied
\cite{LLSSV,DGT}.  The solid curve in Fig.~\ref{fig:powerspectrum} shows this
power spectrum, along with two other power spectra motivated later.  The
corresponding rms mass fluctuations are also shown.

We need to understand how changes to the power spectrum of
density perturbations will affect the abundance of dwarf
galaxies.  We are interested in the abundance of mini-halos of
some given mass that exist within some larger halo of mass
$2\times10^{12}\,M_\odot$, comparable to that
of the Milky Way and the halo mass used by Moore et
al. \cite{Mooetal99}.  The simulations of Moore et al. show that
the number of mini-halos in the Galaxy today is very
close to the number that existed in the proto-galaxy; i.e., no
more than a small fraction are fully disrupted during their
subsequent orbital motion in the Milky Way halo.  Thus, our task 
is simplified: we can calculate the abundance of
mini-halos in the proto-galaxy that later became the Milky Way
halo.  To do so, we use the conditional mass function
\cite{laceycole}, given by
\begin{equation}
     F(>M_{\rm small}) = {\rm erfc}\left[{\delta_{{\rm c}} z_{{\rm f}} 
     \over \sqrt{2 \left[
     \sigma^2(M_{\rm small}) - \sigma^2(M_{\rm big}) \right] }}
     \right] \,.
\label{eq:fraction}
\end{equation}
This equation gives the fraction of the mass in a galactic halo
of total mass $M_{\rm big}$ that was in mini-halos of mass
greater than $M_{\rm small}$ at the redshift $z_{{\rm f}}$ at which the
protogalaxy broke off from the expansion. Here
$\sigma(M)$ is the linear-theory rms fractional mass fluctuation
in spheres of radii that on average enclose a mass $M$, and $\delta_{{\rm c}} = 
1.7$ is the critical threshold for collapse.

\begin{figure}[t]  
\epsfxsize=3.3 in \epsfbox{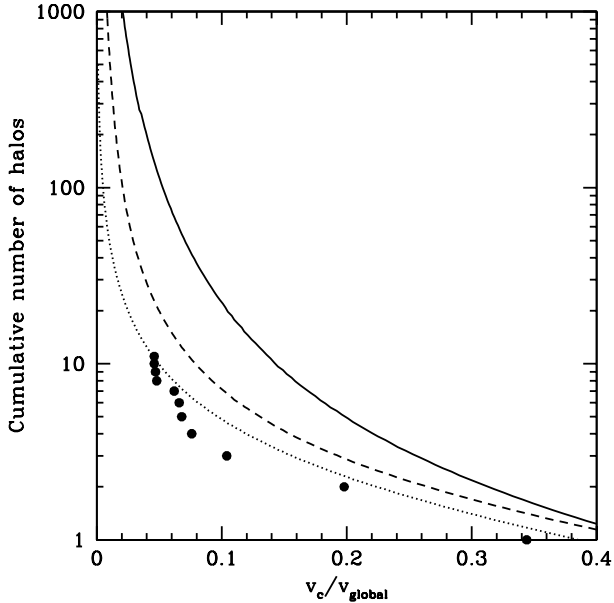}  
\caption{The cumulative number of mini-halos for the power spectra
     shown in Fig.~\protect\ref{fig:powerspectrum} as a function 
     of the circular speed $v_{{\rm c}}$ of the halo divided by the
     circular speed $v_{\rm global}$ of the Galactic halo.  The
     points show the Milky Way satellites.  Compare
     with Fig.~1 in Ref.~\protect\cite{Mooetal99}.
}
\label{fig:halos}
\end{figure}

It is the circular velocity of the
dwarf galaxies, not the mass, which is observed.  We relate the circular speed
$v_{{\rm c}}$ to the mass by assuming the mini-halos underwent collapse
at the protogalaxy-formation epoch.  Then the circular
speed is obtained from $v_{{\rm c}}^3= 10\, M\,G\,
H(z_{{\rm f}})$,
where $G$ is Newton's constant and $H(z)$ is the expansion rate
at the collapse redshift $z$.  
To compare with the numerical results of Moore et al.  \cite{Mooetal99}, we
consider a Galactic halo of mass $M=2\times10^{12} M_\odot$ with a circular
speed of 220 km~sec$^{-1}$.  
Using the conditional mass function, we obtain for the CDM power 
spectrum the cumulative number of halos shown by the solid curve in
Fig.~\ref{fig:halos} (cf. Fig.~1 in Ref.~\cite{Mooetal99}).
Although our calculation of the number of halos has several
shortcomings, the good agreement
with the numerical results shown in Fig.~1 of
Ref.~\cite{Mooetal99} clearly indicates that we are including
the essential physics.  In particular, we reproduce the order-of-magnitude
excess of halos in the theoretical prediction, as compared
to observations, for $v_{{\rm c}}/v_{{\rm global}}$ in the range
0.05 to 0.1.

We now consider how the dwarf-galaxy
abundance changes if the power spectrum is modified on short scales. 
Clearly, if power is reduced on dwarf-galaxy scales ($\lesssim10^{10} \, 
M_\odot$) relative to that
at the Galactic scale ($\sim10^{12} \, M_\odot$), the number of dwarf galaxies 
will be suppressed. To investigate what is required, we first sharply cut off 
the power spectrum at $k = 4.5 h \, {\rm Mpc}^{-1}$, shown by
the dotted curve in  
Fig.~\ref{fig:powerspectrum}; the lower panel shows how the rms mass 
fluctuations are modified. The corresponding prediction for halo
substructure is 
shown as the dotted curve in Fig.~\ref{fig:halos}, and is in
striking agreement with the observations.

However, such a cutoff is purely phenomenological, and much
stronger motivation
is needed. In particular, the question arises as to whether inflation can 
produce such a power spectrum.  The heuristic arguments given
above suggest that 
the power spectrum can undergo a sharp drop on small scales if there
is a discontinuity (or near-discontinuity) in the slope of the potential, and 
this is known as the broken-scale-invariance (BSI) model.
Remarkably, there is 
an exact solution for the power spectrum in this situation, due to Starobinsky 
\cite{Sta92}, even though the slow-roll approximation does not apply. Further, 
this spectrum has a universal form, depending only on the change in the 
derivative across the discontinuity, in the sense that the form
of the spectrum
is preserved if the discontinuity is smoothed out, provided the smoothing is 
over a sufficiently short range of scalar-field values. The underlying reason 
for this is that the field briefly becomes kinetic-energy dominated as it goes 
over the discontinuity, and the precise form of the potential becomes 
irrelevant.

The dashed curve in Fig.~\ref{fig:powerspectrum}
shows this power spectrum for a suitably chosen location and
amplitude of the discontinuity (the amplitude is $p=10$, in the notation of 
Ref.~\cite{LPS}). The universal form features a modest rise, followed by a 
series of oscillations asymptoting to the original spectral shape at a much 
lower amplitude; in this case the power is reduced by a factor $p^2 = 100$ on 
short scales. Such a model represents the fastest possible cutoff in power 
that can be obtained from a single-field inflation model.\footnote{We suspect 
that this is true in multi-field models too, as additional fields will supply 
extra friction, slowing down the fields' evolution and hence
stretching features
in $k$-space.}

This power spectrum gives rise to the dashed curve shown in 
Fig.~\ref{fig:halos}.  This model is almost as successful as the cutoff power 
spectrum; it produces roughly 20 mini-halos with 
$v_{{\rm c}}/v_{\rm global} \gtrsim 0.04$, in much better agreement than the 
original 
power spectrum with the ${\cal O}(10)$ that are observed.
We conclude
that the BSI model provides a promising possibility for
reconciling the predictions with observations.

Our modification to the power spectrum sets in only on very
short scales, and so
does not affect successes of the standard paradigm on much larger scales, such 
as the 
cluster number density or galaxy correlation function. However,
we need to check
that the loss of short-scale power is not inconsistent with
object abundances at
high redshift. The most powerful and direct constraint comes
{}from the abundance
of damped Lyman-alpha systems \cite{Lya}. Observations exist at
redshifts 3 and
4, giving comparable constraints. Redoing the calculation of Ref.~\cite{LLSSV} 
to include updated observations \cite{st2} gives a 95\% confidence lower limit 
of 
\begin{equation}
\sigma(10^{10}\, h^{-1} M_\odot,z=0) > 2.75+h \,,
\end{equation}
in a critical-density Universe, which is (just) satisfied by the BSI model 
(though not by the cutoff power spectrum). 

This narrow escape makes it look as if the model is quite
marginal.  However, we
have only analyzed the critical-density case, and the
damped-Lyman-alpha constraint is much
weaker in a low-density model.  The observational constraint on $\sigma(M)$ at
redshift 4 is almost the same (no more than 10\% lower for $\Omega_0$ values of
interest \cite{LLVW}).  However at redshift 4 suppression of the growth of
perturbations has yet to set in, and the normalization to COBE
(and/or cluster abundance) means $\sigma$ is
higher by a factor $1/\Omega_0$.  With this additional factor all our
models would easily pass the damped Lyman alpha system test were
$\Omega_0 \sim 0.3$.

There are two other key short-scale predictions worth considering. One is the 
power spectrum of perturbations obtained from the Lyman-alpha forest 
\cite{Croetal98,NuHae99}. However, the power spectrum derived in 
Ref.~\cite{Croetal98} does not extend much beyond $k=5h$
Mpc$^{-1}$, which is where our model begins to differ from the usual one. It 
therefore does not appear in 
conflict with the power in our theoretical spectra; indeed, if anything, their 
analyses see less power than expected on short scales. The second test is
whether the model has early enough structure formation to
reionize the Universe
by a redshift 5, as required by the Gunn--Peterson test. Estimates using the 
technique of Ref.~\cite{LL95} would suggest this is marginal in the 
critical-density case. However, once again in the low-density case we benefit 
{}from the much higher early-time normalization of the power spectrum, which 
increases the predicted redshift by a factor $\simeq
1/\Omega_0$. We thus conclude that this BSI model is likely to
be consistent in the observationally-favored low-density flat model,
though a more careful investigation is merited.

The suppression of small-scale power may also help to explain the
discrepancy between the cuspy inner halos observed in simulations and
the smooth halos inferred from observations.  It has been argued
that the steepness of the halo density profile at small radii
depends on how the characteristic density of the mini-halos that
merge to form galactic halos scales with their mass.  If small
halos are denser---as they would be if they collapsed
earlier---then the cusp is steeper \cite{SyeWhi98}.  If
power is suppressed on sub-galactic scales, then all
mini-halos with sub-galactic masses will undergo collapse at
about the same time and thus have similar densities
\cite{NavSte99}.  If so, then halos should be less cuspy than
prior simulations have shown.  However, these heuristic
arguments have been tested numerically for the case of galaxy
clusters \cite{clusters}, where no change to the core structure
was found  when short-scale power was cut off.  More work will
be needed assess the effects on galactic halos.

In conclusion, the arguments
\cite{KlyKraVal99,Mooetal99,KauWhiGui93,NFW} suggesting that the
standard structure-formation paradigm predicts too much
substructure in 
galactic halos have now been refined to an extent that requires that the 
problem be taken very seriously and that radical answers to the
discrepancy be considered. 
We have proposed here that the discrepancy indicates a dramatic
lack of short-scale power in the primordial power spectrum.
While such a sharp feature in the spectrum would not be expected
{\it a priori}, 
it can be obtained from inflation, with the BSI model being
the best available example. This model has already been invoked to introduce 
features on much larger scales of around $100h^{-1}$ Mpc \cite{break,LPS}, but 
the observational motivation for a short-scale break appears stronger. If our 
suggestion is correct, the lack of short-scale power will soon
become apparent in new observations of high-redshift phenomena.

\medskip
We thank Bruno Guiderdoni, Ben Moore, John Peacock, and Paul Steinhardt
for discussions, and we acknowledge the hospitality of the Isaac
Newton Institute where this work was initiated.  MK acknowledges
the support of the DoE, NASA, and the Alfred P. Sloan
Foundation.

\end{document}